\documentclass[12pt,a4paper]{article}
\usepackage{blkarray}
\usepackage{amsfonts}
\usepackage{amsmath}
\usepackage{geometry}
\usepackage{graphicx}
\usepackage{amssymb}
\usepackage{setspace}
\usepackage{geometry}
\providecommand{\U}[1]{\protect\rule{.1in}{.1in}}

\def\fr{\frac}
\def\be{\begin{equation}}
\def\ee{\end{equation}}
\def\ba{\begin{eqnarray}}
\def\ea{\end{eqnarray}}
\def\pa{\partial}

\def\e{\varepsilon}
\def\a{\alpha}
\def\b{\beta}
\def\g{\gamma}
\def\d{\delta}

\def\pt{\phantom{a}}
\def\w{\omega}

\def\na{\nabla}

\begin{document}

\title{How to build a gravity generator}

\author{Diego Marin \thanks{dmarin.math@gmail.com}}

\maketitle

\begin{abstract}
\noindent This article explores a simple way to construct an electric device capable to generate an artificial gravitational field by exploiting a resonance phenomenon.
\end{abstract}

\newpage
\tableofcontents
\newpage

\section{Theoretical foundation}

In \emph{Arrangement Field Theory} all fields fill the adjoint representation of $GL(5,\mathbb{C})\otimes SU(6) \otimes U(1)$. With respect to $GL(5,\mathbb{C})$ we find the following generators:
$${\left(\begin{array}{cc}
0 & (\psi_R)^c \\
\psi_L & 0 \end{array}\right)\atop \text{material fields} \pt(T^a)}\qquad
{\left(\begin{array}{cc}
sl(2,\mathbb{C}) & 0 \\
0 & \mathbf{1}_3 \end{array}\right)\atop \text{gravitational field} \pt(T^b)}\qquad
{\left(\begin{array}{cc}
i\mathbf{1}_2 & 0 \\
0 & -i\mathbf{1}_3 \end{array}\right)\atop \text{gauge fields} \pt(T^c)}$$

\noindent Note that $Tr(T^a T^b T^a T^c) = d^{abac} \neq 0$ so that we have a Feynmann vertex which joins two material fields (i.e. a material current), the gravitational field and a gauge field (for example the electromagnetic field). This fact suggests the existence of a low energy limit in which the following effective lagrangian works:
$$\mathfrak{L} = \a {F^G}_{\a\b} {F^{EM}}^\b_{\pt\,\g} W^{\g\a}$$

\noindent Here $F^G$ is the strength of gravitational field in the GEM approximation, $F^{EM}$ is the strength of electromagnetic field and $W$ is the strength of the material current so defined
\ba W_{\a\b} &=& \pa_\a [\rho \g(v) v_\b] - \pa_\b [\rho \g(v) v_\a]\nonumber \\
             &\cong& \rho [\pa_\a v_\b - \pa_\b v_\a] \nonumber \ea

\noindent with $\g(v)$ Lorentz factor and $v_0 = c$.

\section{Development of the lagrangian}

We use latin letters to label space indices from $1$ to $3$. Then we suppose ${F^G}_{ab}= 0$:
$$\mathfrak{L} = \a [{F^G}_{0 b} {F^{EM}}^b_{\pt\,c} W^{c0}+{F^G}_{a 0} {F^{EM}}^0_{\pt\,c} W^{ca}]$$

\noindent Explicating fields:
\ba W_{c0} &=& -\rho \dot{v}_c \qquad\qquad\qquad  W^{c0} \,=\, \rho \dot{v}_c \nonumber \\
    W^{ca} &=& \rho \e^{cad}(\na \times v)_d \nonumber \\
    {F^{EM}}^{bc} &=& \e^{bcd} B_d \nonumber \\
    {F^{EM}}_{0c} &=& -E_c \qquad\qquad\qquad {F^{EM}}^0_{\pt\,c} \,=\, E_c \nonumber \\
    {F^G}_{0b} &=& -G_b = -\pa_b \Phi \nonumber \ea

\noindent Substituting in $\mathfrak{L}$:
$$\mathfrak{L} = \a' \left[(\pa_b \Phi) \e^{bcd} B_d \dot{v}_c - (\pa_a \Phi) E_c \e^{cad}(\na\times v)_d\right]$$

\noindent with $\a' = -\a \rho$. Varying respect to $\Phi$:
\ba\fr{\d \mathfrak{L}}{\d \Phi} &=& \a' \left[-\pa_b \e^{bcd} B_d \dot{v}_c + \pa_a E_c \e^{cad}(\na\times v)_d\right]\nonumber\\
 &=& \a' \left[\na \cdot (B\times \dot{v}) - \na\cdot(E\times(\na\times v))\right]\nonumber \\
 &=& \a' \na \cdot\left[B\times \dot{v} - E\times(\na\times v)\right]\nonumber\ea

\noindent Thus the Einstein equations gives
$$\na^2 \Phi = \na \cdot G = \a'' \na \cdot\left[B\times \dot{v} - E\times(\na\times v)\right]$$
and then
$$G = \a'' \left[B\times \dot{v} - E\times(\na\times v)\right]$$

\section{Constructing a device}

It's interesting to consider a system with cylindrical symmetry where a fluid rotates with angular velocity $\w$ around an axe parallel to an oscillating magnetic field $B = B_0 cos (\nu t)$. Moreover we consider a cylindrical oscillating electric field $E=E_0 cos (\nu t)$ whose axe coincides with the rotation axe. We take both $B$ and $E$ constant in space. We have
$$ \na\times v \cong -2\w \qquad\qquad \dot{v}= \w^2 \vec{r} + \dot{\w}\times \vec{r}$$

\noindent Hence
\ba G_\theta &=& \a''[\w^2 B r - 2E \w]\nonumber \\
    G_r &=& -\a''B\dot{\w} r \nonumber \ea

\noindent In the canonical form
\ba \dot{\w} &=& \a'''\left[\w^2 B - \fr {2E \w}r \right]\quad \Rightarrow \w_e = \fr{2E}{B r_e} = \fr{2E_0}{B_0 r_e}\nonumber \\
    \dot{r} &=& v_r \quad \Rightarrow v_{r,e} = 0 \nonumber \\
    \dot{v_r} &=& -\a'''B\dot{\w} r \nonumber \\
    &=& -(\a''')^2 B \left[\w^2 B r - 2E \w \right]\quad \Rightarrow \w_e = \fr{2E_0}{B_0 r_e} \nonumber \ea

\noindent with $\a''' = \a''/m$. Now we can expand the equations around $(\w_e, r_e, v_{r,e}=0)$:
\ba \dot{\w} &=& \a'''\left[\w_e^2 B +2\w_e \w B - \fr{2E \w_e}{r_e} - \fr{2E \w}{r_e}+\fr{2E \w_e r}{r_e^2}\right]\nonumber \\
                &=& \a'''\left[\left(\w_e^2 B - \fr{2E \w_e}{r_e}\right)+\left(2\w_e B -\fr{2E}{r_e}\right)\w+\fr{2E \w_e}{r_e^2}r\right]\nonumber \\
                &=& \a''' B \w_e \w + \fr {\a'''B^2\w_e^3}{2E}r\nonumber \\
    \dot{r} &=& v_r \nonumber \\
    \dot{v_r} &=& -(\a''')^2 B \left[\left(\w_e^2 B r_e - 2E \w_e\right)+\left(2\w_e B r_e -2E \right)\w+\w_e^2 B r\right] \nonumber \\
              &=& -2(\a''')^2 E B \w - (\a''')^2 B^2\w_e^2 r\nonumber \ea

\noindent In the matricial form:

$$\left(\begin{array}{c} \dot{\w} \\ \dot{r} \\ \dot{v}_r \end{array}\right)= \left(\begin{array}{ccc} \a'''B\w_e & \fr {\a'''B^2\w_e^3}{2E} & 0 \\ 0 & 0 & 1\\ -2(\a''')^2 E B & -(\a''')^2 B^2\w_e^2 & 0\end{array}\right)\left(\begin{array}{c} \w \\ r \\ v_r \end{array}\right)$$
from which
$$\left(\begin{array}{c} \w \\ r \\ v_r \end{array}\right)= Exp \int dt \left(\begin{array}{ccc} \a'''B\w_e & \fr {\a''' B^2\w_e^3}{2E} & 0 \\ 0 & 0 & 1\\ -2(\a''')^2 E B & -(\a''')^2 B^2\w_e^2 & 0\end{array}\right)\left(\begin{array}{c} \w_0 \\ r_0 \\ v_{r,0} \end{array}\right)$$
$$\left(\begin{array}{c} \w \\ r \\ v_r \end{array}\right)= Exp \left(\begin{array}{ccc} \a'''B_0\w_e\fr{sin(\nu t)}{\nu} & \fr {\a''' B^2_0\w_e^3}{2E_0}\fr{sin(\nu t)}{\nu} & 0 \\ 0 & 0 & t\\ -(\a''')^2 E_0 B_0\left(\fr{sin(2\nu t)}{2\nu}+t\right) & -\fr {(\a''')^2 B_0^2\w_e^2}{2}\left(\fr{sin(2\nu t)}{2\nu}+t\right) & 0\end{array}\right)\left(\begin{array}{c} \w_0 \\ r_0 \\ v_{r,0} \end{array}\right)$$
A case which will appear to be relevant is for $\nu t \ll 1$, i.e. $\nu^{-1} sin(\nu t) \cong t$. The equation reduces as follows:
$$\left(\begin{array}{c} \w \\ r \\ v_r \end{array}\right)= Exp \left(\begin{array}{ccc} \a'''B_0\w_e & \fr {\a'''B^2_0\w_e^3}{2E_0} & 0 \\ 0 & 0 & 1\\ -2(\a''')^2 E_0 B_0 & -(\a''')^2 B_0^2\w_e^2 & 0\end{array}\right)t \left(\begin{array}{c} \w_0 \\ r_0 \\ v_{r,0} \end{array}\right)$$

\noindent The eigenvalues result
$$\g_1 = 0 \qquad\qquad \g_{2,3} = \fr{1\pm i\sqrt 3}{2}\a'''B_0 \w_e$$

\noindent Thus we obtain a resonance frequency
$$\mu = Im\,\g = \fr{\sqrt 3}{2}\a'''B_0 \w_e$$

\noindent Taking $\a''' = t_P q_P/m_P$ we have
$$\mu = 4\cdot 10^{-54}B_0\w_e \,\,s^{-1} = 8\cdot 10^{-54}\fr {E_0}{r_e}\,\,s^{-1}$$

\noindent We can exploit such resonance by using a charged fluid (like a plasma) with charge $Q$ and choosing $\nu = \mu$. In such a way we would have an additional force radially oriented and proportional to $E_0 Q cos(\mu t)$. Resonance frequency is near $0$, so that we can take $E, B$ constant. The only difficult is to work with values of $r$ and $\w$ which touch the equilibrium values:
$$r\w \cong \fr{2E_0}{B_0}$$

\noindent For example
$$E_0 \cong 1\,\,V/m \qquad\qquad B_0 \cong 0,1\,\,T$$
$$r = 1\,\,cm \qquad\qquad \w = 2\,\,KHz = 38'200 \,\,rpm$$

\begin{figure}[h!]
\centering\includegraphics[width=0.8\textwidth ]{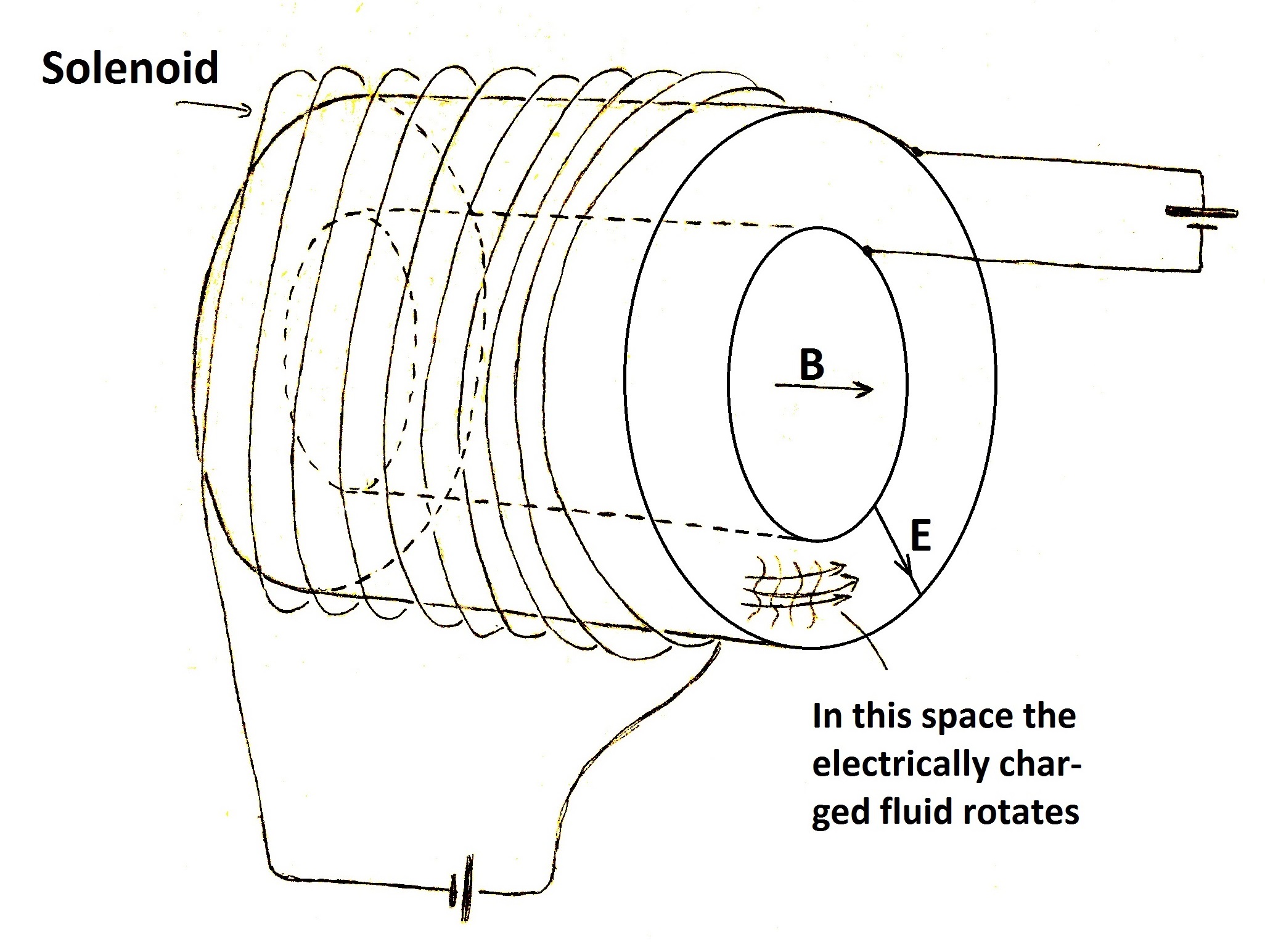}
\label{Device}
\end{figure}










\end{document}